\documentclass[graybox]{svmult}

\usepackage{hyperref}
\usepackage{mathptmx}       % selects Times Roman as basic font
\usepackage{helvet}         % selects Helvetica as sans-serif font
\usepackage{courier}        % selects Courier as typewriter font
\usepackage{type1cm}        % activate if the above 3 fonts are
\usepackage{makeidx}         % allows index generation
\usepackage{graphicx}        % standard LaTeX graphics tool
\usepackage{multicol}        % used for the two-column index
\usepackage[bottom]{footmisc}% places footnotes at page bottom
\usepackage{soul}

\makeindex             % used for the subject index
\newcommand{\beq}{\begin{equation}}
\newcommand{\eeq}{\end{equation}}
\newcommand{\bqa}{\begin{eqnarray}}
\newcommand{\eqa}{\end{eqnarray}}

\newcommand{\erf}[1]{Eq.~(\ref{#1})}

\newcommand{\bra}[1]{\left\langle{#1}\right|}
\newcommand{\ket}[1]{\left|{#1}\right\rangle}

\newcommand{\cu}[1]{\left\{ {#1} \right\}}

\newcommand{\ie}{{\em i.e.}}

\newcommand{\phenom}{{\sc phenomenon}}
\newcommand{\model}{{\sc model}}
\newcommand{\locality}{{\sc locality}}
\newcommand{\sigloc}{{\sc signal-locality}}
\newcommand{\predet}{{\sc predetermination}}
\newcommand{\predict}{{\sc predictability}}
\newcommand{\lc}{{\sc local causality}}
\newcommand{\la}{{\sc local agency}}
\newcommand{\BL}{{\sc Bell-local}}
\newcommand{\ac}{{\sc agent-causation}}
\newcommand{\common}{{\sc common causes}}
\newcommand{\de}{{\sc decorrelating explanation}}%{{\sc causal explanation}} %explanation decorrelates %explaining by decorrelating
\newcommand{\explains}{{\sc explains}}
\newcommand{\Reich}{{\sc Reichenbach}}
\newcommand{\macro}{{\sc macroreality}}
\newcommand{\minkowski}{{\sc Minkowski space-time}}
\newcommand{\order}{{\sc temporal order}}
\newcommand{\arrow}{{\sc causal arrow}}
\newcommand{\cause}{{\sc cause}}
\newcommand{\causes}{{\sc causes}}
\newcommand{\past}{{\sc past}}
\newcommand{\relativity}{{\sc relativistic causality}}
\newcommand{\free}{{\sc free choice}}

\newtheorem{axiom}{Axiom}
\newtheorem{principle}{Principle}
\newtheorem{postulate}{Postulate}

\definecolor{ngreen}{rgb}{0.2,0.7,0.2}%161
\definecolor{nred}{rgb}{0.9,0.1,0}%711&900
\definecolor{nblue}{rgb}{0.1,0.2,0.8}
\newcommand{\blk}{\color{black}}

\newcommand{\hmw}[1]{}%{{\color{ngreen} \small \bf [[{#1}]]}}
\newcommand{\egc}[1]{}%{{\color{blue} \small \bf [[{#1}]]}}

\begin{document}

\title*{{\em Causarum Investigatio} and the Two Bell's Theorems of John Bell}
%\titlerunning{Causation and the Two Bell's Theorems} 
%for an abbreviated version of
\author{Howard M. Wiseman and Eric G. Cavalcanti}
\institute{Howard M. Wiseman \at Centre for Quantum Dynamics, Griffith University, Brisbane, Queensland 4111, Australia. \\ email: H.Wiseman@griffith.edu.au \and 
Eric G. Cavalcanti \at School of Physics, The University of Sydney, Sydney, NSW 2006, Australia. \\ email: e.cavalcanti@physics.usyd.edu.au}
\maketitle

\abstract{``Bell's theorem'' can refer to two different theorems that John Bell proved, the first in 1964 and the second in 1976. 
His 1964 theorem is the incompatibility  of quantum phenomena with the joint assumptions  of \locality\ and \predet. 
His 1976 theorem is their incompatibility with the single property of \lc.  
This is contrary to Bell's own later assertions, that his 1964 theorem began with the assumption of \lc, 
even if not by that name. 
Although the two Bell's theorems are logically equivalent, their assumptions are not. Hence, the earlier and later 
theorems suggest quite different conclusions,  embraced by operationalists and realists, respectively. 
 The key issue is whether \locality\ or \lc\ is the appropriate notion emanating from \relativity, 
 and this rests on one's basic notion of causation. For operationalists the appropriate notion is what is here called 
 the Principle of \ac, while for realists it is \Reich's Principle of common cause. By breaking down the latter
 into even more basic Postulates, it is possible to obtain a version of Bell's theorem in which each  
 camp could reject one assumption, happy that the remaining assumptions reflect its {\em weltanschauung}.  
 Formulating Bell's theorem in terms of causation 
is fruitful not just for attempting to reconcile the two camps, but also for better describing the ontology of different 
quantum interpretations and for more deeply understanding the implications of Bell's marvellous work. }

\section{Motivation}
\label{sec:motiv}
%$\hat{\rm P\!I}$, $\hat{\rm \Pi\hspace{-1.6ex}P}$ 

The work presented here grew from my\footnote{This first section is written in first person by one of us (Wiseman), 
who spoke at the Quantum [Un]speakables conference. The other of us (Cavalcanti) has been  
a long-time co-worker and correspondent with Wiseman on Bell's theorem. In particular, 
since the conference, their discussions have convinced Wiseman of a better way to formulate 
the causal assumptions in Bell's theorem, and this is reflected in the latter parts of the paper, 
and in its authorship.} observation, over the years, but particularly
at a quantum foundations conference in 2013 (see Ref.~\cite{blog13}), that different `camps' of physicists and philosophers 
have a different understanding of what Bell's theorem actually is, and a different understanding of the 
words (in particular `locality') often used in stating it. As a consequence they often talk (or shout) 
past one another, and come no closer to understanding each other's perspective. I have friends in 
both camps, and I would like to think that they are all reasonable people who should be able to come 
to terms that allow the pros and cons of different interpretations of Bell's theorem to be discussed 
calmly. 

\begin{figure}[t]
\includegraphics[width=\textwidth]{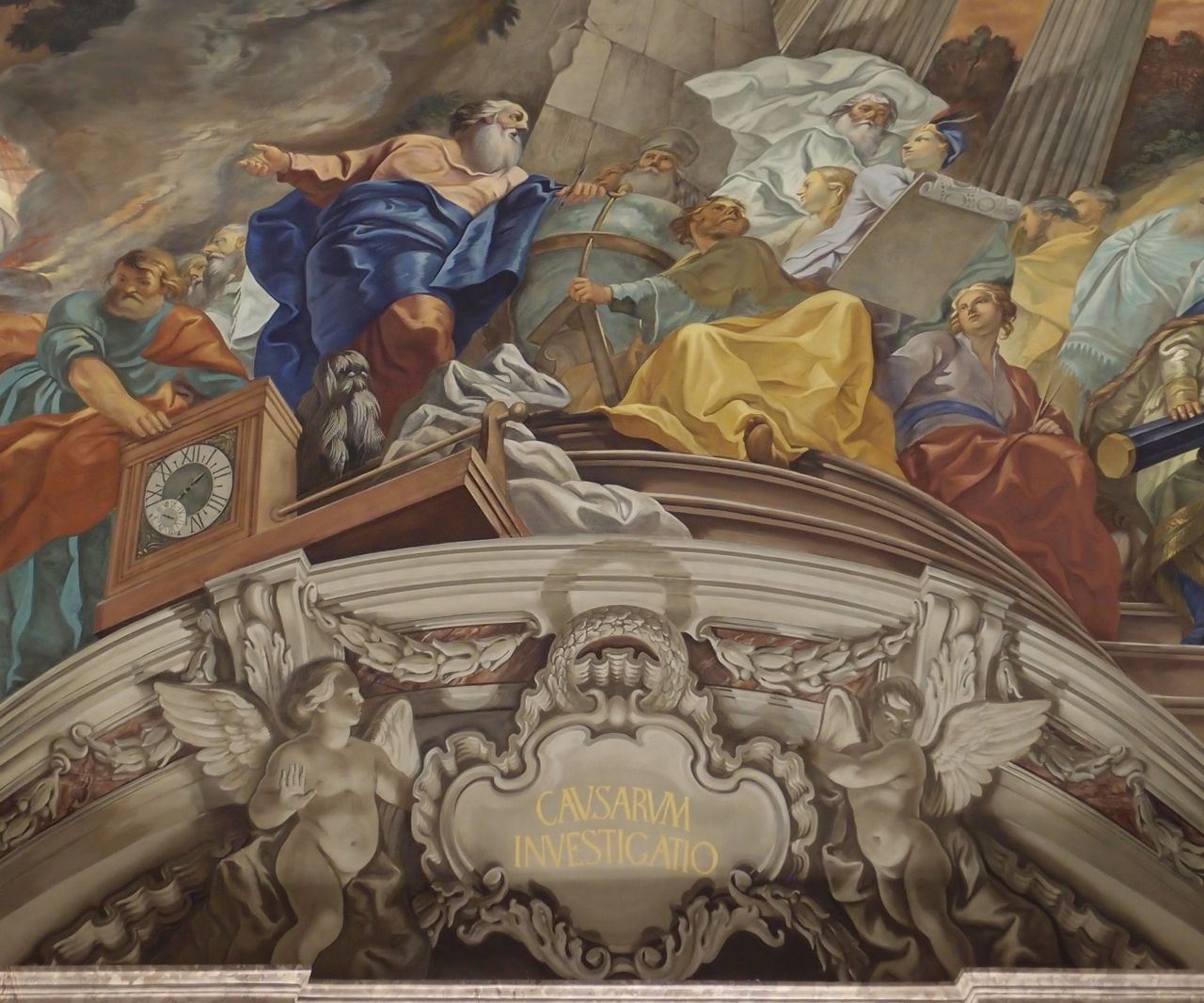}
\caption{Scene from the ceiling of the {\em Festsaal} of the  Austrian {\em Akademie der Wi\ss{}enschaften}}
\label{fig:CI}       % Give a unique label
\end{figure}

More recently, my thinking has evolved beyond this original motivation,
as a consequence of the 50th anniversary of Bell's {\em annus mirabilis}\footnote{It was the year he wrote his 
review of hidden variables (HVs)~\cite{Bel66}, by misfortune not published until 1966, in which he dismissed 
von Neumann's anti-HV proof, gave the first proof of the necessity of contextuality for deterministic HV theories, 
and raised the question of the necessity of nonlocality, which he immediately answered in the positive in his 1964 paper~\cite{Bel64}.}, which led to three invitations 
to present on the topic: for a special issue of J. Phys. A \cite{Wis14b}, for an opinion piece in Nature \cite{Wis14a}, 
and for the Quantum [Un]speakables conference (and thus to a fourth invitation, for these Proceedings). 
These challenged me to think more deeply about what lay behind the different positions of the 
two camps, and how this could lead to a deeper understanding of Bell's theorem. The key to my reconsideration is to be found, 
conveniently, on the the magnificent ceiling of the {\em Festsaal} of the  Austrian {\em Akademie der Wi\ss{}enschaften}, 
host to Alain Aspect's public lecture at the Quantum [Un]speakables conference. There, one finds the role of natural 
philosophy defined as {\em causarum investigatio} (Fig:~\ref{fig:CI}). % (the investigation of causes). 
The investigation of causes, or, more particularly, of notions of causation, 
has proven to be a very fruitful way to analyse Bell's theorem, and the disagreements over it~\cite{vFr82,WooSpe15,CavLal14,Wis14b,Wis14a}. 
%In addition to my own~\cite{Wis14a,Wis14b}, there are some other notable recent papers adopting this perspective on 
%Bell's theorem~\cite{WooSpe12,CavLal13}, but with a somewhat different motivation. 

While this chapter, Ref.~\cite{Wis14a}, and Ref.~\cite{Wis14b}, each has 
independent material, they all share some core material, which can 
be summarized as follows. Bell actually proved two theorems (Sec.~\ref{sec:2Bells}). The 1964 one~\cite{Bel64} is that 
\begin{quote}
[some of] the statistical predictions of quantum mechanics are incompatible with separable [\ie~local] predetermination.
\end{quote}
This involves two assumptions: ``separability or locality'' and the ``predetermination [of] the result of an 
individual measurement''~\cite{Bel64}. The 
%and quantum phenomena are incompatible with the conjunction of locality (L) and predetermined outcomes (D); and the 
1976 one~\cite{Bel76} is that 
\begin{quote}
quantum mechanics is not embeddable in a locally causal theory
\end{quote}
and involves a single assumption, ``local causality''~\cite{Bel76}. 
%quantum phenomena are incompatible with local causality (LC). 
Although each theorem is a corollary of the other (Sec.~\ref{sec:Fine}), 
they are embraced by different intellectual camps, whom I will call operationalists and realists respectively (Sec.~\ref{sec:2camps}). 
The latter, however, deny that there are really two different theorems, claiming that in his 1964 paper Bell 
used `locality'  to mean \lc, and that from it he derived 
\predet\footnote{In earlier, albeit recent, publications~\cite{Wis14a,Wis14b} I used the 
term `determinism' for Bell's second assumption in 1964. However, Bell did not actually use this word in 1964, 
and it is useful to reserve it for a different notion~\cite{WisCav15}. %(stronger in some ways; weaker in others) as
%will be discussed elsewhere~\cite{WisCav15}. 
Note the use here of small-capitals for terms with a precise technical meaning, 
even when the relevant definition is given only at a later point in the chapter.}  
rather than assuming it. I have argued in depth~\cite{Wis14b,WisRie15} that this is a misrepresentation of what Bell proved in 1964.  

Whether \locality\ or \lc\ is the appropriate notion emanating from Einstein's Principle of relativity 
rests on one's underlying concept of causation. Operationalists and realists implicitly hold to quite different notions 
of causation and it is fruitful to make this explicit (Sec.~\ref{sec:basics}). These last points lead on to the notable new material in 
this chapter: a form of Bell's theorem that could be acceptable to both camps (Sec.~\ref{sec:randr}), and discussion of the many other 
advantages of formulating Bell's theorem in terms of causation (Sec.~\ref{sec:conc}). 

\section{The situation Bell considered}\label{sec:Belsit}

\begin{figure}[t]
\includegraphics[width=\textwidth]{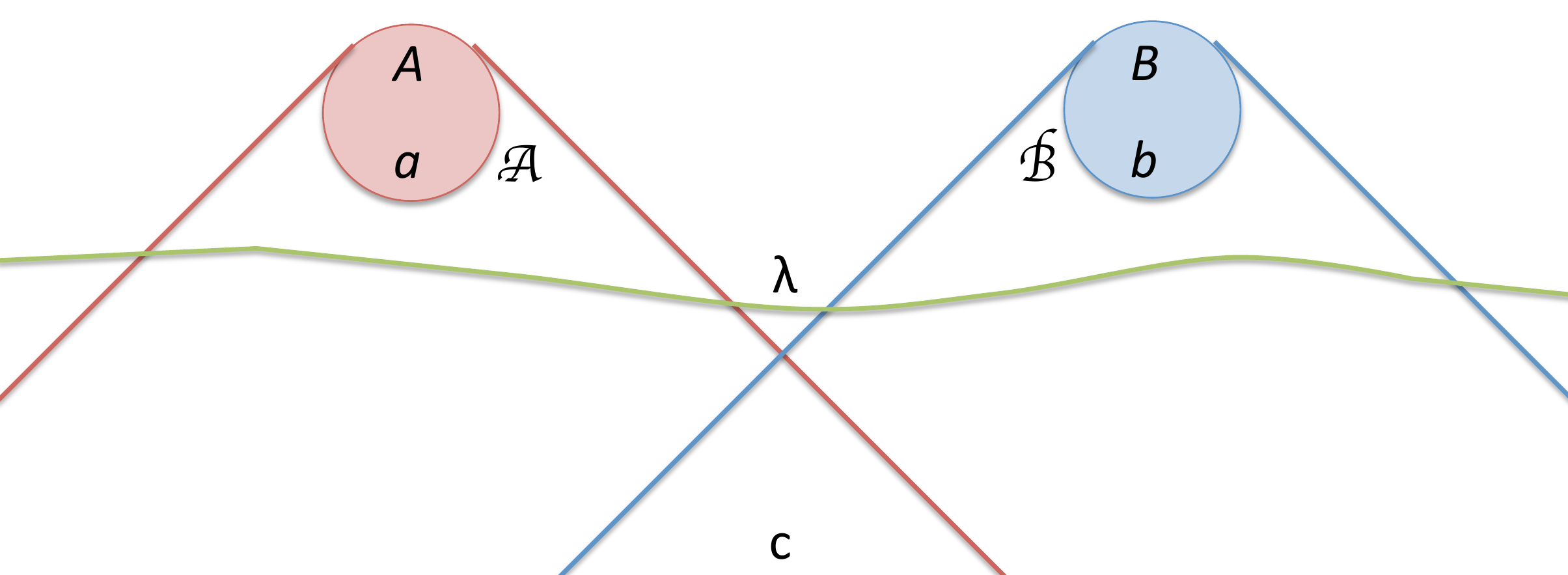}
\caption{Minkowski diagram for the scenario Bell introduced in 1964. Time is the vertical axis, space horizontal, and the diagonal lines indicate the boundaries of light-cones.}
\label{fig:MD}       % Give a unique label
\end{figure}

The experimental situation Bell considered is shown as a Minkowski (space-time) diagram in Fig.~\ref{fig:MD}. 
This is closely based on such diagrams which Bell used in Refs.~\cite{Bel76,Bel90b}, for example. 
He did not use such a diagram in 1964, and only briefly referred to relativistic concepts, but the diagrams 
he used in his definitive paper on the subject~\cite{Bel90b} are applicable to, and even 
use the same notation as, his original 1964 paper. %Since this chapter is concerned with 
%contrasting (albeit briefly, compared to Ref.~\cite{Wis14b}) the operationalist and realist camps, 
%which draw their inspiration from ... theorems respectively
%\blk{both of Bell's theorems (1964 and 1976),  } 
It is convenient 
to use something close to the form of diagram which Bell finally settled upon~\cite{Bel90b}, as 
it allows the assumptions from both 
those theorems (\locality\ and \predet\ in 1964; \lc\ in 1976) to be stated naturally. 
Considering more general distributions of events in space-time lead to different conclusions about 
which assumptions can be stated naturally; see Sec.~\ref{sec:conc} and Ref.~\cite{WisCav15}. 

The experiment involves two observers, Alice and Bob, with spatially separated laboratories, 
where they perform experiments that are independent, but may have correlated outcomes. 
This allows them to perform experiments in space-like-separated regions of space-time, labelled 
${\cal A}$ and ${\cal B}$ respectively. In her region, Alice makes a free choice $a$ of setting for her 
measurement, which yields an outcome $A$, and Bob likewise, {\em mutatis mutandis}. 
In the overlap of the past light-cones of ${\cal A}$ and ${\cal B}$ is a preparation event $c$ which is 
necessary to produce the correlations between the outcomes. Even for a fixed preparation $c$, the 
ability of Alice and Bob to make free choices, and the existence of multiple possible outcomes, 
gives rise to a \phenom\ described by the relative frequencies\footnote{\label{foot:frequencies}Note that here we are not making a commitment to frequentism, but rather simply recognising that real experimental data are relative frequencies.}
\beq
\{ f(A,B| a, b, c) : A, B, a, b \}.
\eeq
The existence of correlations in the outcomes in a physics experiment is typically `explained' 
by stepping away from operationalist language, in a manner such as this: the event $c$ prepares 
a correlated pair of particles, one of which goes to ${\cal A}$ 
and the other to ${\cal B}$, each at subluminal speed. 

Stepping even further away from operationalism, we follow Bell in allowing for the possibility of 
non-observable (or ``hidden'') variables 
in the formalism, introducing variables $\lambda$, defined at a time --- \ie~on a
space-like hypersurface (SLH) --- before 
$a$ and $b$ but after $c$, that could describe these particles, or anything else of 
relevance\footnote{More generally~\cite{Bel90b}, one could sandwich these variables 
between two suitable SLHs, but the above formulation will suffice.}.  
Note in particular that $\lambda$ could represent a pure {\em quantum state} $\ket{\lambda}$, 
since this is a mathematical object defined on a SLH. If the source $c$ reliably produces this particular 
pure state then this $\ket{\lambda}$ would contain no more information than is already given by specifying $c$. 
But this is an idealization, 
and in general $\ket{\lambda}$ would have some distribution given $c$, and so is just like any other 
hidden variable. In general, we say that a \model\ (which Bell~\cite{Bel64} called a `theory') $\theta$ 
for the above \phenom, comprises a probability distribution for hidden variables, 
$P_\theta(\lambda|c)$, and further probabilities $P_\theta(A,B|a,b,c,\lambda)$, such that\footnote{Given footnote \ref{foot:frequencies}, this equation is not to be read as a strict equality, but as carrying the same meaning as that of any probabilistic prediction.}
\beq \label{def:model}
\sum_\lambda P_\theta(A,B|a,b, c, \lambda) P_\theta(\lambda|c) = f(A,B|a,b,c).
\eeq

Note that by considering a \model\ with non-trivial dependence on $\lambda$, one 
is {\em not} presuming \predet\ of outcomes:
\begin{definition}[\predet]  \label{def:predet}
$\forall\ A,a,B,b,\lambda, \
  P_\theta(A,B|a, b, c, \lambda) \in \cu{0,1}$.
\end{definition}
(See the next section for the source of this terminology.) An example of a hidden variable \model\ that is stochastic 
(\ie~that violates \predet) is the one mentioned above, where $c$ prepares a mixed quantum state $\rho_c$ and $\lambda$ is taken to define a pure 
quantum state $\ket{\lambda}$ such that $\rho_c = \sum_\lambda  P_\theta(\lambda|c) \ket{\lambda}\bra{\lambda}$. 

Note also that \predet\ should not be confused with the stronger and purely operational notion of \predict~\cite{CavWis12}:~
\begin{definition}[\predict]  
$\forall\ A,a,B,b,    \   f(A,B|a, b, c) \in \cu{0,1}$.
\end{definition}
\blk While we have defined \predict\ in terms of the experimental frequencies $ f(A,B|a, b, c)$, for a given \model\  
it is of course possible to determine whether the \phenom\ it is supposed to describe has this property, via \erf{def:model}. 
The point is that a \model\ may satisfy \predet\ even though the \phenom\ does not satisfy \predict.

\section{The Two Bell's Theorem of John Bell}\label{sec:2Bells}

In this section we present the two theorems, in chronological order, and then discuss the relationship between them.

\subsection{Bell's 1964 theorem}  \label{sec:Bel64}

In Bell's 1964 paper he states what he has proven most clearly in his Conclusion (Section VI): 
\begin{quote}
{\em In a theory in which parameters} are added to quantum mechanics to {\em determine the results of individual measurements}, without changing the statistical predictions, there must be a mechanism whereby {\em the setting of one measuring device  can influence the reading of another instrument, however remote}.
\end{quote}
Here the italics, added by us, emphasize the two assumptions that lead to a contradiction with the statistical predictions of quantum mechanics; 
the second assumption is stated in the negative, since its negation follows if one holds to the first assumption. These two assumptions 
are stated positively, with equal status, in the immediate preceding sentence, at the end of his Section V:
\begin{quote}
for at least one \ldots\ state \ldots\ the statistical predictions of quantum mechanics are incompatible with separable predetermination
\end{quote}
as quoted in Sec.~\ref{sec:motiv} above. As was noted there, Bell did not distinguish separability from locality, and he is explicit that \predet\ means 
``predetermination [of] the result of an individual measurement''. Thus his theorem is:
\begin{theorem}[Bell-1964] 
There exist quantum {\sc phenomena} for which there is no \model\ satisfying \predet\ and \locality.
\end{theorem} 
Although this was not the case in 1964, the quantum phenomena relevant to Bell's theorem have 
long since been verified experimentally, albeit with a few challenging loopholes~\cite{Wis14a,Loopholes2014}.

It is presumably uncontroversial to understand \predet\ as per Definition~\ref{def:predet} already given above. 
The meaning of \locality\ is more controversial, to say the least (compare Refs.~\cite{Nor06,Nor15} with Refs.~\cite{Wis14b,WisRie15}). 
However, by our reading, Bell is quite clear:
\begin{quote}
the requirement of locality [is] more precisely that the result of a measurement on one system be unaffected by operations on a distant system
\end{quote}
This is of course the positive form of the final notion in the first quote in this section, and Bell states the same assumption (the irrelevance of Alice's measurement choice to Bob's outcome) twice more 
in the paper. Although he successfully applies the notion of locality only to theories with predetermined outcomes, 
he introduces it prior to making the assumption of predetermination\footnote{\label{foot:EPRpara}He introduces it in the first paragraph of the paper proper, which serves as motivation for the formulation of the assumptions he will use. There, Bell unfortunately misapplies his notion, in attempting to derive 
predetermination via EPR-Bohm-correlations and locality. See Refs.~\cite{Wis14b,WisRie15} for discussions  of the irrelevance of this flawed paragraph to Bell's 1964 theorem (\ie\ the ``result to be proved'', as he calls it, which he does indeed prove).}. Thus it seems that we should adopt 
a definition that accords with his words and applies to probabilistic theories:
%\begin{definition}[\locality] \label{def:local}
%$$\forall\ a,B,b,c,\lambda,  \ 
% \exists \ P^\beta_\theta(B|b, c, \lambda) :   P_\theta(B|a, b, c, \lambda) = P^\beta_\theta(B|b, c, \lambda),$$
% \end{definition}
%\egc{Why the "there exists" -- is that to avoid defining lambda as a "sufficient specification"? And what is the meaning of the superscript beta -- do we really need it?}\hmw{No, because in a nonlocal theory $P^\beta_\theta(B|b, c, \lambda)$ is not necessarily a well-defined distribution. 
%It would be like asking for $P^\beta_\theta(B|c, \lambda)$ with no conditioning $b$. Do you prefer this?} 
\begin{definition}[\locality] \label{def:local}
$$\forall\ a,B,b,\lambda,  \ 
P_\theta(B|a, b, c, \lambda) = P_\theta(B| b, c, \lambda),$$ 
 \end{definition} 
and likewise for Alice's result, which will remain unstated in similar definitions below. 
(The existence of the function $P_\theta(B| b, c, \lambda)$ is also implicit here, and in similar 
definitions below.) This definition of \locality\ was also the one adopted by Jarrett in 1984~\cite{Jar84}. 
In the same year Shimony~\cite{Shi84} coined the phrase ``parameter 
independence'' for the same concept, to emphasize that it required only that Bob's outcome be statistically independent of Alice's 
setting, a controllable `parameter'. We prefer to follow the terminology of Jarrett and (in our reading) Bell.

Note that \locality\ is not the same~\cite{CavWis12} as the purely operational notion of
 \begin{definition}[\sigloc] \label{def:sigloc}
$$\forall\ a,B,b,  \   f(B|a, b, c) = f(B|b, c),$$ 
  \end{definition}
 the violation of which has never been observed, and, most physicists think, 
never will be observed. However, for a strictly operational \model\ (that is, one that makes no use 
of hidden variables, be they pure quantum states or other objects) there is no distinction 
between \locality\ and \sigloc. Thus, operational quantum mechanics, involving only 
preparations, settings, and outcomes, satisfies \locality. In fact, \locality\ is satisfied even 
in the non-operational quantum theory discussed above, where $c$ prepares $\rho_c$ 
but one assumes %considers a particular ensemble $\sum_\lambda P(\lambda|c) \ket{\lambda}\bra{\lambda}= \rho_c$, 
%and says 
that, conditional on the hidden variable $\lambda$, the probabilities of outcomes 
are determined by $\ket{\lambda}$. %This is not operational because there are 
%infinitely many such ensembles and no way to verify the `true' one.  
%$\lambda$ represents a pure quantum state, which is not strictly operational if $c$ is taken 
%to prepare a mixed state,\egc{One could say that, strictly operationally, $c$ defines a mixed state but not a particular ensemble of $\lambda$.} satisfies \locality. 
In both cases, the reduced quantum state on Bob's side,
which is mixed in general, defines probabilities for Bob's outcomes, for any measurement he makes, 
which are independent of Alice's choice of measurement. Moreover, introducing 
a realist narrative, involving instantaneous wave-function collapse, makes no difference 
to the fact that \locality\ is satisfied, as a mathematical statement about the probability distributions in 
the theory. Even spontaneous 
collapse theories such as the celebrated GRW model~\cite{GRW86} respect \locality\ 
(to the extent that they respect \sigloc).

 \subsection{Bell's 1976 theorem} \label{sec:Bel76}
 
 More than a decade after his 1964 paper, Bell reformulated his theorem in a way that he would 
cleave to, in essence, for the rest of his life. Building on his own work from 1971~\cite{Bel71} and that 
of Clauser and Horne from 1974~\cite{CH74}, he introduced a new notion:
\begin{definition}[\lc]  \label{def:lc}
$$  \forall\ A,a,B,b,\lambda,  \ 
  P_\theta(B|A, a, b, c, \lambda) = P_\theta(B|b, c, \lambda).
$$
\end{definition}
Strictly speaking (and Bell was strict about this~\cite{Bel76}), the above is a consequence of the broader Principle of \lc\ 
(Principle~\ref{pr:lc} in Sec.~\ref{sec:repr} below) when applied to the specific set up of Fig.~\ref{fig:MD}.  
Note the crucial difference from \locality\ in that \lc\ requires Bob's outcome to be statistically independent of Alice's outcome 
as well as her setting. This, Bell argued, is a reasonable `localistic' notion for a theory in 
which $\lambda$ provides a complete description 
of the relevant state of affairs prior to the measurements being performed.

As quoted in Sec.~\ref{sec:motiv}, Bell proved in 1976 that 
\begin{quote}
quantum mechanics is {\it not} embeddable in a locally causal theory as formulated [above]. %above
\end{quote}
In other words, he proved a theorem involving only a single assumption: 
\begin{theorem}[Bell-1976] 
There exist quantum {\sc phenomena} for which there is no \model\ satisfying \lc. 
\end{theorem}
Unfortunately for our purposes, having invented a new concept with a new name, 
Bell immediately became indiscriminate once more, using ``local causality'' and `locality' 
interchangeably\footnote{Indeed, by 1981~\cite{Bel81} he was implying that by `locality' he 
had always meant \lc. This historical revisionism is perfectly understandable, 
and probably unconscious --- a plausible unfolding of the localistic intuition Bell had in 1964 is \lc, since this 
would have worked, where \locality\ failed, in Bell's attempt to reproduce the EPR argument (see footnote~\ref{foot:EPRpara}).}.
However, in his most mature treatment of the subject \cite{Bel90b}, Bell  unequivocally showed his preference 
for the term ``local causality''~\cite{Wis14b}, and in following suit we respect Bell's final will. %In the below, 
%I will also use the term Bell-(non)locality for (the violation of) \lc, terms 

\subsection{A Fine Distinction} \label{sec:Fine}

Bell's 1976 theorem implies Bell's 1964 theorem. This is because, as is easy to see, 
any \model\ $\theta$ satisfying \locality\ and \predet\ satisfies \lc. The converse of that last clause  
is not  true; there are theories satisfying \lc\ that do not satisfy \locality\ and \predet. 
Orthodox quantum mechanics with separable states is one example. Nevertheless, 
the converse of the first sentence of this subsection {\em is} true. That is, Bell's 1964 theorem 
implies Bell's 1976 theorem. This is because of the following:  
%any demonstration of ``Bell nonlocality'' (violation of \lc) also violates the joint assumptions of \predet\ plus \locality. 
\begin{theorem}[Fine-1982] 
For any \phenom,  there exists a \model\ $\theta$ satisfying \lc\ if and only if there exists a \model\ $\theta'$ satisfying \predet\ and \locality. 
\end{theorem}
Although we have given the credit here to Fine~\cite{Fin82}, this result was known, in essence, by Bell even in 1971~\cite{Bel71}; 
see Ref.~\cite{Wis14b} for details. For this reason, it is useful shorthand to introduce the well-known term 
Bell-local for describing the type of \model\ that satisfies the broader assumptions of Bell's 1976 theorem. That is, %\begin{definition}[\BL]  \label{def:BellLocal} A \phenom\ is \BL\ if and only if there exists a {\sc theory} for it satisfying \lc: 
%$$f(A,B|a,b,c) = \sum_\lambda P^\alpha_\theta(A|a, c, \lambda)P^\beta_\theta(B|b, c, \lambda) P(\lambda|c).   
%$$
\begin{definition}[\BL]  \label{def:BellLocal} A \model\ $\theta$ %for a \phenom\ 
is \BL\ if and only if %it has this form:
\beq \label{eq:BellLocal}
\forall\ A,a,B,b,  \
P_\theta(A,B|a,b,c) = \sum_\lambda P_\theta(A|a, c, \lambda)P_\theta(B|b, c, \lambda) P_\theta(\lambda|c).   
\eeq
\end{definition}
%Here, for brevity, the existential quantifiers for the local distributions are omitted.  

If the two Bell's theorems are logically equivalent, why should we bother to distinguish them?
The answer is that the two different forms appeal to two different camps of scholars, and indeed these 
two camps often recognise only the one form that they favour. The broad term `scholars' here is
deliberately chosen to cover the increasing range of disciplines --- including (at least) physics, philosophy, and information science --- 
interested in Bell's theorem. But it is important to note that the division into two different camps does not sharply follow these 
disciplinary divisions. Of course there are many more than two attitudes towards Bell's theorem. Nevertheless, 
the most common attitudes can be broadly grouped within the two camps, called here 
operationalist and realist. 

\section{The Two Camps}\label{sec:2camps}

The realist camp~\cite{DGZ92,Mau94,Nor06,Mau14} has the following credo: \begin{quotation}
Bell's theorem uses only one assumption: local causality (or `locality' as we usually call it for short). 
This is the only reasonable way to apply the principle of relativity for statistical theories. It is essentially what EPR assumed in 1935. They showed that operational quantum mechanics is nonlocal, and Bell showed in 1964 that adding hidden variables cannot solve the problem. 
Experiments violating a Bell inequality thus leave us with no option: the principle of relativity is false.  The world is nonlocal. 
\end{quotation}
The operationalist camp~\cite{vFr82,Mermin93,NieChu00,ZukBru14}, on the other hand,  could be caricatured by the following: \begin{quotation}
Bell's theorem uses two assumptions. The first assumption is locality. This is essentially the same as signal locality, which is all the principle of relativity implies, but also applies to hidden variable theories. Orthodox quantum mechanics respects locality. The second assumption is something else which has been variously called realism,  predetermination, determinism,  objectivity,  classicality,  counter-factual-definiteness, and causality (perhaps with slightly different formulations). Clearly it is the second assumption that we should abandon, whatever we call it. Locality is here to stay.
\end{quotation}

Why do the two camps come to such contrary conclusions? Partly it is just a difference in terminology: realists 
use `locality' to mean \lc, while operationalists use it to mean \locality. But the deeper question is why they disagree
about which is the `correct' way to apply the principle of relativity. Bell well explains the motivations of the realist camp~\cite{Bel90b}: 
\begin{quotation}
The obvious definition of ``local causality" does not work in quantum mechanics, and this cannot be attributed to the ``incompleteness" of that theory.  \ldots\

Do we then have to fall back on ``no signalling faster than light" as the expression of the fundamental causal structure of contemporary theoretical physics? That is hard for me to accept. For one thing we have \emph{lost the idea that correlations can be explained} \ldots. More importantly, the \emph{``no signalling" notion rests on concepts which are desperately vague} \ldots. The assertion that ``we cannot signal faster than light" immediately provokes the question:  
``Who do we think {\it we} are?'' 
%{\it We} who can make ``measurements"? {\it We} who can manipulate ``external fields"? 
%{\it We} who can ``signal" at all, even if not faster than light? Do {\it we} include
%chemists, or only physicists; plants or only animals; pocket calculators, or only mainframe computers?
\end{quotation} 
Here the italics, added by us (except for the `{\em we}' in the final question), emphasize the two key tenets of the realist camp: that correlations should be explained, 
and that anthropocentric notions such as `signalling' should play no fundamental role. 

An operationalist, however, could 
well claim to know well enough who `{\em we}' are, and point out that that statements about what we may, or may not, be able to do 
are very useful, for example in informational security proofs. From an operationalist perspective, moreover, explanations, in the sense Bell means, might be regarded as superfluous.  These   
differences %in perspective 
between realists and operationalists hark back to Einstein's 1919 distinction between constructive theories
and principle theories~\cite{Ein19}. %http://todayinsci.com/E/Einstein_Albert/EinsteinAlbert-MyTheory.htm
 But a more precise understanding of the origin of the disagreement is possible by breaking down the assumptions 
 used in Bell's theorem to a more basic level (Sec.~\ref{sec:basics}). As we will see, this is also the way towards enabling the two camps
 to discuss Bell's theorem using the same terms, agreeing on what combinations of assumptions 
 it implies to be impossible\footnote{This is always assuming that a loophole-free test, expected soon~\cite{Wis14a,Loopholes2014} will 
 not turn up any surprises.}, even while disagreeing on which assumption is most implausible (Sec.~\ref{sec:randr}). 
 
 \section{Back to Basics} \label{sec:basics}
 
Here we make a fresh start, aiming to base Bell's theorem on notions more comprehensive and 
more fundamental  than those defined earlier in this chapter with reference to the particular scenario of Sec.~\ref{sec:Belsit}. 
Those earlier notions are temporarily abandoned, but they will gradually be reintroduced 
and their relation to the deeper concepts indicated. 

To begin, we introduce some axioms. In calling them `axioms' we are not implying that they are unquestionable, 
only that we will not question their necessity in the formulations of Bell's theorem below. Without further ado:  
\begin{axiom}[\macro] \label{axiom:macro}\begin{svgraybox}
An event observed by any observer is a real single event, and not `relative' to anything or anyone.  \end{svgraybox}
\end{axiom}
This rules out consideration of the ``relative state''~\cite{Eve57} or ``many worlds''~\cite{DeWGra73} interpretation, 
as well as the extreme subjectivism of the `QBist' interpretation~\cite{FucMerSch14}.~
\begin{axiom}[\minkowski] \label{axiom:STC} \begin{svgraybox}
Concepts such as space-like separation, light-cones, and foliations of SLHs are well defined in ordinary laboratory situations. \end{svgraybox}
\end{axiom}
This rules out short-cuts through space (`wormholes') between distant locations~\cite{Holland95}. 
There is actually no need to restrict to flat space-time; any time-orientable Lorentzian space-time manifold 
will do, but the above terminology is more straight-forward. 
\begin{axiom}[\order] \label{axiom:order} \begin{svgraybox}
For any event $A$, there is a SLH containing $A$ that separates events in the \past\ of $A$ from events that have $A$ in their \past. %; equivalently, any \effect
 \end{svgraybox} 
\end{axiom} 
Note the term \past\ %, \effect, and \future\ 
is not, in this axiom, to be understood as having definite meaning; 
the font used is meant to alert the reader to this fact. In particular, there is no implication that all events on one side of the SLH are in the \past\ of $A$  and all events on the other side in the future of $A$ (\ie\ having $A$ in their \past). For example, \order\ is satisfied if we take \past\ to 
mean the past light-cone, which defines only a partial ordering of events (that is, for 
some pairs of events, neither is in the \past\ of the other). To define an almost-total ordering of events 
(that is, such that for almost every pair of events, one of them is in the \past\ of the other) one would need to define \past\ via a fixed foliation of SLHs. 
\begin{axiom}[\arrow] \label{axiom:arrow} \begin{svgraybox}
 Any \cause\ of an event is in its \past. %; equivalently, any \effect\ of an event must be in its \future.
 \end{svgraybox}
%Causation acts only from the  past   to the future. 
%with $c$ in the past of $a$, and $a$ in the past of $A$ {\it etc.}  
\end{axiom} 
This axiom, together with axioms~\ref{axiom:STC} and \ref{axiom:order}, implies a causal structure describable by a directed acyclic graph, as is standard in modelling of causation~\cite{WooSpe15}.  
It rules out retrocausal approaches to Bell's theorem such as in Refs.~\cite{Peg80,Price08}. Note that, like \past\ above, \cause\ here does not yet have any precise meaning. The meaning of these concepts will become more defined as more assumptions are added. 

In moving now from axioms to postulates, we list assumptions that are more likely to be questioned, or at least that were questioned, in some form, relatively early in the literature on Bell's theorem. Indeed, if one accepts the Axioms then one must (modulo the remaining experimental loopholes) 
reject one of the postulates below, as we will show in Sec.~\ref{sec:randr}. %, but a realist would make a different choice from an operationalist. 
The first postulate (which was listed as an Axiom in Ref~\cite{Wis14b}) begins the process of adding 
meaning to `\cause': 
\begin{postulate}[\free] \label{po:fc} \begin{svgraybox}
A freely chosen action has no relevant {\sc causes}. %\blu{in common with any relevant causes of the physical systems under study.} 
\end{svgraybox}
\end{postulate} 
Here, `relevant' means ``in common with, or among, the other events under study.'' 
This postulate is not meant to indicate a philosophical commitment to Cartesian dualism, or a religious 
commitment to Pelagianism, although  it is possibly incompatible with 
Augustinian predestination\footnote{{\em Les passions de l'\^a{}me} (1649), {\em De natura} (415), and {\em  De natura et gratia} (415) respectively.}. Here, it serves to rule out (when combined with 
other principles) what %Postulate~\ref{po:common}
Bell called `superdeterminism'~\cite{Bel90b}. The `super' in `superdeterminism' indicates that 
free choices would not just have causes as a matter of principle (as believed by those who hold to the 
`determinist' philosophy of free will). Rather, they would of necessity have causes that are correlated in a very 
particular way with external physical variables that affect the outcomes of measurements.  Although this 
viewpoint has at least one prominent scientific proponent~\cite{tHooft02}, our personal view is that 
Postulate~\ref{po:fc} is as unquestionable as any of the Axioms above; see footnote~\ref{foot:alien} below. 
%to derive theorems one requires strictly worded assumptions like Postulate~\ref{po:fc}.  
\begin{postulate}[\relativity]\label{po:relativity}\begin{svgraybox}
The \past\ is the past light-cone. % In the Axioms, ... of \arrow
%, and the \future\ is to be understood as the future light-cone. 
\end{svgraybox}
\end{postulate} 
Note that a term like ``past light-cone'' {\em is} to be understood as having definite meaning, 
from the Axiom of \minkowski. 
\begin{postulate}[\common]\label{po:common} \begin{svgraybox}
If two sets of events ${\cal A}$ and ${\cal B}$ are correlated, and no event in either is a \cause\ of any event in the other, then 
they have a set of common {\sc causes} ${\cal C}$ that \explains\ the correlation. \end{svgraybox}
\end{postulate}
Here,  {\em common} {\sc causes} means events that are {\sc causes} of at least one event in 
${\cal A}$ and at least one event in ${\cal B}$. It is important to note 
that this postulate is not the same as \Reich's Principle of common cause~\cite{Rei56}. 
Rather, following Ref.~\cite{CavLal14} (although with some differences in details) 
we have deliberately split Reichenbach's celebrated principle into the above Postulate 
of \common, and the below
 \begin{postulate}[\de]\label{po:de} \begin{svgraybox}
A set of {\sc causes} ${\cal C}$, common to two set of events ${\cal A}$ and ${\cal B}$, \explains\ a correlation between ${\cal A}$ and ${\cal B}$  only if conditioning on ${\cal C}$ eliminates the correlation. \end{svgraybox}
\end{postulate}
% \begin{postulate}[\de]\label{po:de} \begin{svgraybox}
%A set of events ${\cal C}$ \explains\ a correlation between two other sets of events ${\cal A}$ and ${\cal B}$  only if conditioning on ${\cal C}$ eliminates the correlation. \end{svgraybox}
%\end{postulate}
As Reichenbach said~\cite{Rei56},  ``When we say that the common cause explains the [correlation], we refer \ldots\ to 
the fact that relative to the cause the events $A$ and $B$ are mutually independent.''

In the above principles we referred always to events, but for statistical concepts such as correlation it is more common for physicists
to think in terms of variables. In such cases we will be loose with notation and terminology, and allow, for instance, $A$ to stand for 
the outcome that Alice gets (a variable) as well as the event that Alice's outcome takes the value $A$. 
%In such examples, because $(A,a,c)$ a single event.

\subsection{Realist Principles} \label{sec:repr}

The axioms and postulates above form (as we will see in Sec.~\ref{sec:randr}) a nice set in that they are sufficient to 
enforce {\sc Bell-locality}, and, with one exception, all necessary. (The one exception is \order, 
which is not needed if one assumes \relativity.)  
%Hmmm. 
%Level 1: macro, Minkowski, arrow
%Level 2: order, free, cc
%Level 3: relativity(=>cc), Reich(=>in cc) 
However they do not correspond to the principles stated in Bell's two theorems, 
and hence do not obviously connect to the two camps. Thus we will develop principles that do 
relate to the two camps (starting with realists) and show their relation to the postulates 
above. 

First, since the explanation of correlations is a realist tenet, realists hold to \Reich's Principle, 
which we state here explicitly, 
 \begin{principle}[\Reich]\label{pr:reich} 
If two sets of events ${\cal A}$ and ${\cal B}$ are correlated, and no event in either is a \cause\ of any event in the other, then 
they have a set of common {\sc causes} ${\cal C}$, such that 
conditioning on ${\cal C}$ eliminates the correlation. 
\end{principle}
For realists, this is the point of causation --- to explain correlations. However we made 
that a separate assumption, as discussed above, and as captured by this: 
\begin{lemma} The postulates of \common\ and \de\ imply \Reich's Principle.
\end{lemma}

For realists, the role of \Reich's principle is in this: 
\begin{lemma} \Reich's Principle and the Postulate of \relativity\ imply the Principle of \lc,
\end{lemma}
if we define the last generally, basically as Bell did in 1976, as 
\begin{principle}[\lc] \label{pr:lc}
If two space-like separated sets of events ${\cal A}$ and ${\cal B}$ are correlated, then 
%\st{they have a set of causes} 
there is a set of events ${\cal C}$ in their common Minkowski 
past such that conditioning on ${\cal C}$ eliminates the correlation.
\end{principle}
Here the {\em common} Minkowski past means the intersection of the union of past light-cones of events in ${\cal A}$ with 
the union of past light-cones of events in ${\cal B}$. 

The reader may well ask how this Principle of \lc\ relates to the definition~\ref{def:lc}. 
%\blk Returning now to the derivation of \lc\ as per definition~\ref{def:lc}, 
%from the Principle of \lc, 
Take the two sets of events in Principle~\ref{pr:lc} to be ${\cal A} = (A,a)$ and ${\cal B} = (B,b)$. 
%We can include the relevant settings as part of the outcome event because a record of these 
%settings can be assumed to exist at the time and place the outcome is obtained. 
Thus if \lc\ is satisfied 
there must exist a set of {\sc causes} ${\cal C}$ in their common Minkowski 
past such that 
$P(A,a,B,b|{\cal C})=P(A,a|{\cal C})P(B,b|{\cal C})$. 
\blk Thus, $P(B,b|A, a, {\cal C}) = P(B,b|{\cal C})$, and $P(B|A, a, b, {\cal C}) = P(B|b, {\cal C})$. 
Replacing ${\cal C}$ by Bell's variables $c$ and $\lambda$,\footnote{\blk The alert reader will have noticed 
some sleight of hand. In Bell's 1976 paper, where he introduced \lc, 
$\lambda$ denoted all events in the intersection of the past light-cones. But in Ref.~\cite{Bel90b} Bell took the $\lambda$s 
to be defined between two SLHs, and the limit when these become one corresponds to the situation he 
considered in 1964, where the $\lambda$s were ``initial values of the [relevant] variables at some suitable instant.'' 
As in Fig.~\ref{fig:MD}, that suitable {\em instant} (SLH) may not even cross the intersection of the past light-cones.   
The resolution is that %the {\em principle} of \lc, as enunciated above, is more broadly applicable than the 
%definition \ref{def:lc}. The former refers to correlations between any space-like-separated events, not just to 
%observable outcomes. If, 
since, by assumption, the variables $\cu{\lambda}$ are the only relevant ones, 
they must carry the information that was present in the common causes ${\cal C}$ in the common Minkowski 
past \blk
of $A$ and $B$. This they can without falling foul of \lc\ (in the sense of Principle~\ref{pr:lc}) because there is a part of the SLH that is in the future light-cone of the events in ${\cal C}$, but in the past 
light-cone of $A$, and likewise for $B$.} 
a sufficient specification of the causes by assumption, gives definition~\ref{def:lc}. 
%P(B|bC)=P(Bb|C)/P(b|C)=P(Bb|C)P(C)/P(b,C)=P(Bb|C)P(C)/P(C|b)P(b)

\blk This is not, however, sufficient 
to derive a contradiction with quantum phenomena. It was sufficient in Sec.~\ref{sec:2Bells} 
because there we were working within the framework of our definition of a \model\ in \erf{def:model}. 
Here we want a more principled derivation of the condition of {\sc Bell-locality}, in 
\erf{eq:BellLocal}. For this it is necessary to use an additional assumption related to the Postulate of \free. 
Specifically, we combine this postulate with other postulates: 
\begin{lemma} \label{lem:la}
The Postulates of \relativity\ and \common\ and \free\ imply the Principle of \la,  
\end{lemma}
where we have defined (for terminological reasons to be explained elsewhere~\cite{WisCav15})  
%\egc{So if we change FREE CHOICE as I've suggested, I suppose we would need to change LOCAL AGENCY to "..., among the variables under study".}
\begin{principle}[\la] \label{pr:la}
The only relevant events correlated with a free choice are in its future light cone. 
\end{principle}
This assumption was not stated precisely by Bell in 1976, who said only 
``It has been assumed that the settings of instruments are in some sense 
free variables \ldots.''\footnote{Bell was immediately criticised for the vagueness of this statement 
(and for what followed, some of which was not sufficient for his purpose) by 
Shimony, Horne, and Clauser~\cite{ShiHorCla76}. The immediacy was, according 
to Clauser~\cite{Cla02}, because the latter two authors had originally drafted their 1974 paper~\cite{CH74} 
using the above Principle of \lc\ (or something like it), but Shimony pointed out to them that this was not sufficient 
to derive \BL{\sc ity} without an extra assumption relating to free choice. As a consequence they retreated 
from such a principled formulation of \lc\ to the more specific definition~\ref{def:lc},  
which they said characterized ``objective local theories''~\cite{CH74}, enabling  
a ``less general and more plausible''~\cite{ShiHorCla76} assumption (than \la, for example) 
relating to free choice. Clauser and Horne~\cite{CH74} deserve credit for first (as far as we are aware) 
\blk discussing, in their footnote 13, 
\blk the need for independence of the hidden variables $\lambda$ from the free choices 
$a$ and $b$ which is implicit in \erf{def:model}.  They note that to justify that assumption 
 one has to rule out the ``possibility'' that  
``Systems originate within the intersection of the backward light cones of both analyzers and the source 
\ldots [which] effect [{\em sic.}] both the experimenters' selections of analyzer orientations and the emissions from the source.'' 
}. However,  according to Bell in 1977~\cite{Bel77} what he meant by this, 
in the context of theories satisfying \lc, was that ``the values of such variables have implications 
only in their future light cones,'' in other words, the above Principle of \la. 

It is worth noting, however, that Bell does not actually need a Principle as strong as this, 
but rather only (as he says two sentences later) that ``In particular they have no implications for 
the \ldots\ variables in \ldots\ [their] backward light cones.''  This weaker assumption follows 
(using Axioms~\ref{axiom:order} and \ref{axiom:arrow} without \relativity) 
from the following %Principle of \ac:  
\begin{principle}[\ac] \label{pr:ac}
If a set of relevant events ${\cal A}$ is correlated with a freely chosen action, then that 
action is a \cause\ of at least one event in ${\cal A}$. 
\end{principle}   
This can be seen as follows. 
Assume that $a$ is correlated with some event $A$. From this Principle of \ac, $a$ must be a cause of $A$. From 
Axiom~\ref{axiom:arrow}, $a$ must be in the \past~of $A$. From Axiom~\ref{axiom:order}, this means there 
must be a SLH separating $a$ and $A$. Thus, as claimed, $A$ cannot be in the past light-cone of $a$. 
%(Principle~\ref{pr:ac}, in the next subsection). 
If one is making the assumption of \lc, then the Axioms need not be used, and one can directly conclude 
from Principle~\ref{pr:ac} that $A$ must be in the future light-cone of $a$. That is, the Principles of \ac\ 
and \la\ become equivalent.

\begin{figure}[t]
\begin{center}
\includegraphics[width=0.84\textwidth]{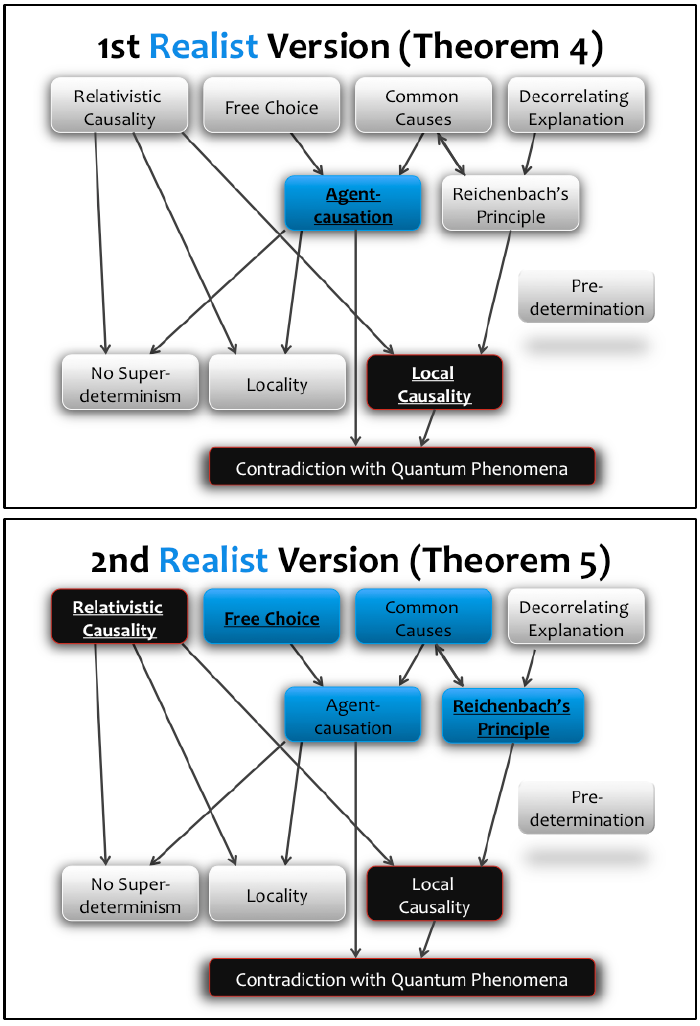}
\vspace{-0.2cm}
\end{center}
\caption{\blk Two realist versions of Bell's Theorem, \ref{thm:4} and \ref{thm:5}. 
The coloured (shaded) boxes are concepts used in the formulation in question, or upheld in the philosophy 
of the camp (here, realists) in question. The black boxes are concepts used in the formulation in question, but  
rejected in the philosophy of the camp in question. The white (pale) boxes are concepts not used in the 
formulation in question and ignored in the philosophy of the camp in question. \underline{\bf Underlined bold font} is used for 
the fundamental assumptions (2, 3, or 4 in number) of the formulation in question. The arrows indicate
that a particular concept holds if all concepts pointing to it (plus the Axioms) hold.\blk}
\label{fig:theorem45}       % Give a unique label
\end{figure}

\blk Now, to obtain \erf{eq:BellLocal} we must consider $P(A,B|a,b,c)$, which can be written as 
$\sum_{\lambda} P(A,B|a,b,\lambda,c)P(\lambda|a,b,c).$ With the 
location of events in Fig.~\ref{fig:MD}, if the events in ${\cal C}$ are in the 
common Minkowski past of ${\cal A}$ and ${\cal B}$, 
they are not in the future light-cone of either $a$ or $b$. Thus from the principle of \la,
%We will see in Sec.~\ref{sec:oppr} that the postulates 
%we have already used, plus that of \free, imply therefore that $a$ and $b$ cannot be correlated with ${\cal C}$. 
%With this condition, 
$P(\lambda|a,b,c)=P(\lambda|c)$. Using \lc\ as above then gives \erf{eq:BellLocal}. 
 Thus we have a more principled version of Bell's 1976 theorem: 
\begin{theorem}[Bell-1976, in principles]  \label{thm:4}
Quantum {\sc phenomena} violate the conjunction of  
Axioms~\ref{axiom:macro}--\ref{axiom:arrow}, the Principle of \lc, and the Principle of \la\  (or \ac). 
\end{theorem}
From the lemmata in this section (noting that \Reich's Principle implies the Principle of \common, and applying this in 
Lemma~\ref{lem:la}), we can also formulate this theorem using the more 
fundamental postulates as follows:
\begin{theorem}[Bell-1976, in deeper principles]  \label{thm:5}
Quantum {\sc phenomena} violate the conjunction of  
Axioms~\ref{axiom:macro}--\ref{axiom:arrow}, the Postulate of \free, the Postulate of \relativity, and \Reich's Principle. 
\end{theorem}
\blk These two theorems, and the relationship between them are illustrated in Fig.~\ref{fig:theorem45}.\blk
%We admit that this is not the most elegant explanation, but it is the 
%scenario Bell considered in 
%The set of events $\lambda$ on the SLH in question, can be a common cause for events in 
%$\alpha$ and $\beta$ because 
% It implies that any subset of $\lambda$ only in the past light-cone of $\alpha$, 

\subsection{Operationalist Principles} \label{sec:oppr}

As discussed in Sec.~\ref{sec:2camps}, a key difference between realists and operationalists 
is that the former seek to {\em explain} correlations while the latter do not (or, at least, not in the same sense). 
We have shown how realists can enshrine this goal in causal terms by \Reich's Principle. Another key 
difference is that operationalists are happy to put the actors in centre-stage. 
Thus, operationalists should be happy to accept  
%In this section we begin by formulating 
a notion of causation which is not about explaining all correlations, and which is agent-centric: the Principle of \ac\ 
(Principle~\ref{pr:ac} above). 
%\begin{principle}[\ac] \label{pr:ac}
%If a set of relevant events ${\cal A}$ is correlated with a freely chosen action, then that 
%action is a \cause\ of at least one event in ${\cal A}$. 
%\end{principle}   

Just as combining \Reich's Principle with the Postulate of \relativity\ gives \lc\ (the realists' favoured localist notion), 
here we have: 
\begin{lemma} The Principle of \ac\ plus the Postulate of \relativity\ implies the Principle of \locality,
\end{lemma}
where we can provide the following principled version of \locality: 
\begin{principle}[\locality] 
The probability of an observable event $A$ is unchanged by conditioning on a space-like-separated free choice $b$, 
even if it is already conditioned on other events not in the future light-cone of $b$. 
%\rep{any of the {\sc causes} of $A$.}
%what is wrong with "causes of A"? Answer: how do we know what the causes of A are? B could be a cause of A! Only makes 
%sense if one is using RC as well, which is not what we want. Want it to stand alone. 
%"uncorrelated with $b$ not work because B could be uncorrelated with b even though caused by. 
%The point is, uncorrelated given what?
%Ooooh, another point: how do we enforce $a$ being independent of $b$. Not actually ruled out by sd. Maybe not matter??
\end{principle}
%Has to imply P(A|abc,lambda) = P(A|bc,lambda). Tick.
%Has to give factorizability in a general sense when combined with predetermination and no-sd. I guess.
%Has to be obeyed by Orthodox QM. 
%Has to give signal locality in a general sense with suitable minimal modification. 
This Lemma can be demonstrated as follows. 
%\st{Define ${\cal A}$ to be an observable event $A$ and its {\sc causes}. By Postulate~\ref{po:relativity}, these 
%{\sc causes} must be in the past light-cone of $A$. Thus if $A$ is space-like-separated from $b$, then so 
%are all events in ${\cal A}$, and by Postulate~\ref{po:relativity} cannot have $b$ as a \cause. 
%If $b$ is a freely chosen action, then by Principle~\ref{pr:ac}, ${\cal A}$ cannot be correlated with $b$. 
%Hence the probability of $A$ conditional on any of its \causes\ is independent of $b$.}
Define ${\cal A}$ to be $A$ and other events not in the future light-cone of $b$. 
By Postulate~\ref{po:relativity}, none of the events in ${\cal A}$ can have $b$ as a \cause. 
If $b$ is a freely chosen action, then by Principle~\ref{pr:ac}, ${\cal A}$ cannot be correlated with $b$. 
Hence the probability of $A$, even conditional on other events in ${\cal A}$, is independent of $b$. 

Note that \locality\ is by no means the strongest principle that can be derived from \ac\ and \relativity;
one can also show 
\begin{lemma} The Principle of \ac\ plus the Postulate of \relativity\ implies the Principle of \la,
\end{lemma}
as defined in the preceding section (Principle~\ref{pr:la}). We use \locality\ rather than \la\ 
to remain faithful to Bell's original concept. However, we do also use these principles to 
derive another principle that was implicit for Bell in 1964:
\begin{lemma}\label{lem:nosup}
The Principle of \ac, and the Postulate of \relativity\ imply the Principle of {\sc no superdeterminism},
\end{lemma} 
where this last is: 
\begin{principle}[{\sc no superdeterminism}]  
Any set of events on a SLH is uncorrelated with any set of freely chosen actions subsequent to that SLH.
%A freely chosen action is not correlated with any events in its past. 
\end{principle} 
%Here ``in its past'' does not necessarily mean in its past light-cone. It could in fact be replaced 
%by ``not in its future light-cone'', 
The name of this Principle is taken from Bell~\cite{Bel90b}, and its form chosen 
in keeping  with the assumption of \predet\ below. 
% of in that case we would prefer a different name for the principle. 
%Since 
%the lemmata of this section also imply that we can rule out {\sc superdeterminism} from 
%\relativity, \Reich, and \free. This is the result that we appealed to in Sec.~\ref{sec:repr}. 

\begin{figure}[t]
\begin{center}
\includegraphics[width=0.84\textwidth]{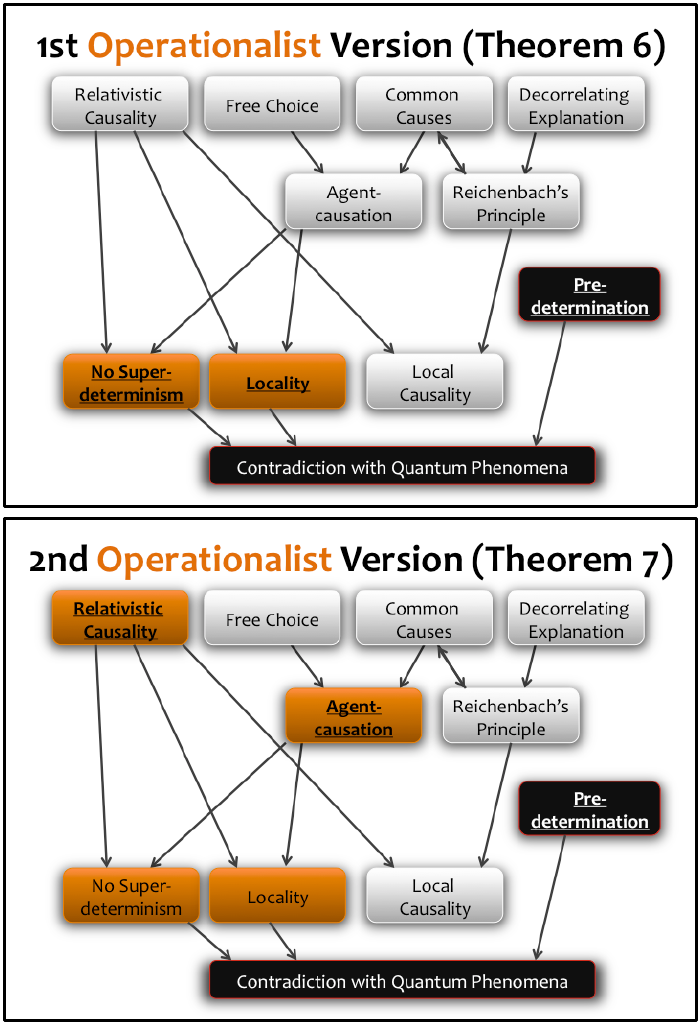}
\vspace{-0.2cm}
\end{center}
\caption{\blk Two operationalist versions of Bell's Theorem, \ref{thm:6} and \ref{thm:7}. For explanatory details, see Fig.~\ref{fig:theorem45}.}
\label{fig:theorem67}       % Give a unique label
\end{figure}

Unlike \lc, \locality\ is not sufficient to make a theory \BL, even with the Principle of {\sc no superdeterminism}. 
 We require an additional principle, which, based on Bell's 1964 paper, we formulate as: 
\begin{principle}[\predet]  
For any observable event $A$, and any SLH $S$  prior to it,  
$A$ has  \causes\ 
%Put in \causes\ here because this, I think, would allow one to define Einstein's 
%``independent real situations'' (realism?) in a future publication as \predet\ plus \relativity.} \rep{prior to}
on $S$, which, possibly in conjunction with free choices 
subsequent to $S$, determine $A$. 
\end{principle} 
%\st{Here we say ``prior to that SLH'' rather than ``on that SLH'' because it makes no real difference and seems more natural.}
%\hmw{It does make a difference if we want to use \predet\ plus \relativity\ to define IRS.}
Applying this to Bell's scenario, choose the SLH to be prior to both $a$ and $b$, 
so that  Bob's outcome is a function $B(a,b,c,\lambda)$. By the Principle of \locality, the dependence on $a$ 
must be trivial, because it is space-like-separated from $B$. Finally, by the Principle of {\sc no superdeterminism}, 
the probability of $\lambda$ cannot depend on $a$ and $b$. Doing the same for $A$, 
\blk and following the same argument as in Sec.~\ref{sec:repr} we see that {\sc Bell-locality} is obeyed. Thus, 
\begin{theorem}[Bell-1964, in principles] \label{thm:6}
Quantum {\sc phenomena} violate the conjunction of  
Axioms~\ref{axiom:macro}--\ref{axiom:arrow} and the Principles of {\sc no superdeterminism}, \locality, and \predet. 
\end{theorem}
From the lemmata in this section, we can also formulate this theorem using more fundamental, 
and not more numerous, principles, as follows: 
\begin{theorem}[Bell-1964, in deeper principles] \label{thm:7}
Quantum {\sc phenomena} violate the conjunction of  
Axioms~\ref{axiom:macro}--\ref{axiom:arrow}, the Postulate of \relativity, and the Principles of \ac\ and \predet. 
\end{theorem}
This has the advantage of not using the term `locality', which realists generally use with a different meaning,  
as discussed above. These two theorems, and the relationship between them are illustrated in Fig.~\ref{fig:theorem67}. \blk

%?? the causes of any observable event can be taken to be events prior to any given SLH, plus free choices 
%subsequent to that SLH.
%
%predetermination: any observable event is determined by its causes.
%
%rel cause: causes in past light-cone

%So: could have: no superdet, locality, and predetermination.   

%comes from rel causality and:
%
%??? : prob observable event, is unchanged if conditioned on a free choice that is not its cause, 
%even if already conditioned on any of its causes.

\section{Reactions and Reconciliation} \label{sec:randr}

Let us now review the forms of Bell's theorem favoured by the two camps, as formulated in 
terms of causal principles deeper than those used by Bell in either 1964 or 1976. 
In all of the below (as in the related figures), we assume the usual
\begin{enumerate}
 \item Axiom  of \macro,
 \item Axiom of \minkowski,
 \item Axiom of \order,
 \item Axiom of \arrow.
 \end{enumerate}
 
 First, the realist's version makes the additional assumptions of 
\begin{itemize}
\item The Postulate of \free,
\item The Postulate of \relativity, 
\item \Reich's Principle. 
\end{itemize}
Of these three, realists reject \relativity, leaving the question of what defines 
\past\ %and \future\ 
in the Axiom of \arrow\ open to further physical investigation. 
Operationalists, on the other hand, would reject (or indeed have rejected~\cite{vFr82}) \Reich's Principle. 
But the latter have reason to be unhappy with the remaining list, because the remaining Axioms and Postulates have  no empirical consequences. They do not imply even \sigloc, for example. 
Thus the operationalists would be left 
saying that they privilege the Postulate of \relativity\ over \Reich's Principle 
even though they cannot point to any consequences of believing in the former. 
For this reason the operationalists would, presumably, object to the list of options offered by this form of 
 Bell's theorem. 

Next we consider the operationalist's version, which assumes instead
\begin{itemize}
\item The Postulate of \relativity,
\item The Principle of \ac,
\item The Principle of \predet. 
\end{itemize}
Of these three, operationalists reject \predet, embracing the idea of intrinsic randomness.  
Realists, on the other hand, would presumably still reject the Postulate of \relativity. 
But the latter have reason not to be happy with the list of options 
offered here. The reason is that \predet\ is just too easy to reject. Even though many of the theories realists take seriously, 
such as de Broglie--Bohm mechanics~\cite{deB56,Boh52}, satisfy \predet, it is not a necessary feature of realist theories, 
and realists would not want to be characterized as basing their rejection of \relativity\ on a belief in \predet. Realists 
do not reject the Principle of \ac, because it is implied by \Reich's Principle and \free\ (this is just a weaker version of
 Lemma~\ref{lem:ac} below). 
But it is 
the fact that \ac\ is weaker than \Reich\ that is the problem, they might say, because it 
entails the addition of the third assumption (\predet) which is clearly too strong, 
and so too easy to reject.  Thus realists do, in fact~\cite{Mau94}, object to a form of Bell's 
theorem listing \predet\ as a fundamental assumption. %these types of options.  
%offered by this form of Bell's theorem. 

\begin{figure}[t]
\begin{center}
\includegraphics[width=0.84\textwidth]{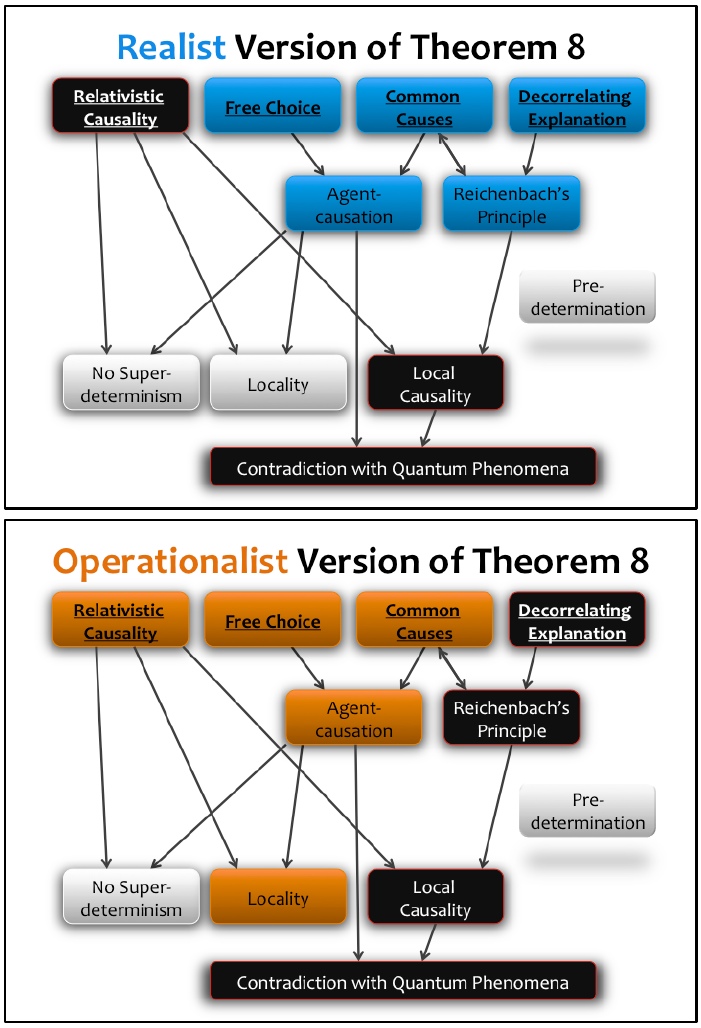}
\vspace{-0.2cm}
\end{center}
\caption{\blk The realist and operationalist interpretations of our `reconcilation' version of Bell's Theorem, \ref{thm:rec}. For explanatory details, see Fig.~\ref{fig:theorem45}.\blk}
\label{fig:theorem8p}       % Give a unique label
\end{figure}

It would seem that the two camps are still at an impasse. But by returning to the 
Postulates of Sec.~\ref{sec:basics} we can bridge the gap. From results already discussed, it is apparent that 
{\sc Bell-locality} can be derived from the Axioms plus
\begin{itemize}
\item The Postulate of \free, 
\item The Postulate of \relativity,
\item The Postulate of \common,
\item The Postulate of \de. 
\end{itemize}
For realists, little has changed from their preferred formulation, and they will as before, reject \relativity. 
But operationalists now should be happy to reject \de, and keep the other postulates. The reason is that 
\begin{lemma} \label{lem:ac}
The Postulates of \free\ and \common\ imply the Principle of \ac.
\end{lemma}
%Although here presented as an alternative notion of causation to \Reich's Principle, we note that 
%\begin{lemma} \label{lem:ac}
%The Postulate of \free, plus \Reich's Principle implies the Principle of \ac.
%\end{lemma}
This can be seen as follows. Say a set of events ${\cal A}$ is 
correlated with a free choice $a$. Now by Postulate~\ref{po:fc}, $a$ has no relevant {\sc causes}.
Thus by %Principle~\ref{pr:reich} 
Postulate~\ref{po:common}, the only option is that $a$ is a \cause\ of at least one event in ${\cal A}$, as in Principle~\ref{pr:ac}. 
%This is just a stronger version of Lemma~\ref{lem:ac}. 
Thus operationalists still have their agent-centric notion of causation, which operational 
quantum mechanics respects, even with the addition of 
the Postulate of \relativity. Some following this direction believe that a replacement for the Principle 
of \de\ is open to further physical investigation~\cite{LeiSpe13}.  The attitude each camp  
would be expected to take to this theorem is illustrated in Fig.~\ref{fig:theorem8p}. \blk

\section{Conclusion}\label{sec:conc} 
%Other advantages of a causal formulation}

After thorough investigation, we suggest that the cause of the disagreement between 
operationalists and realists over Bell's theorem is a disagreement over causes. 
This leads the two camps to favour Bell's 1964 theorem and Bell's 1976 theorem respectively, 
because the localistic notions they employ relate to notions of agent-causation and explanatory
causation respectively. However, by breaking down notions into more fundamental postulates, we 
could formulate what we believe is the best version of Bell's theorem, for the purposes of reconciling the two camps: 
\begin{svgraybox}
\vspace{-2ex}\begin{theorem}[Bell-reconciliation]\label{thm:rec}
Quantum {\sc phenomena} violate the conjunction of  
Axioms~\ref{axiom:macro}--\ref{axiom:arrow} and the Postulates of \free, \relativity, \common, and \de.
\end{theorem}
\end{svgraybox}

This formulation avoids contentious words like `locality', and allows each camp to 
reject one assumption (\de\ and \relativity, respectively), knowing 
that the remaining assumptions reflect its philosophical position. 
Of course even if the two camps do agree upon a single form of Bell's theorem, 
their disagreement about which assumption to reject is still a substantive disagreement. 
But at least they could discuss the merits of their {\em weltanschauungen} using a common vocabulary, 
and so avoid talking past one another, or (the next stage) interminable arguments about what 
terms like `locality' should mean. If, in the end, they just agree to disagree, that would still be a 
great improvement over the present state of affairs~\cite{blog13,Mau14,Wer14,Mau14a,Wer14a}.

Another advantage of the above formulation is that \relativity\ better reflects the ontology of different 
quantum interpretations than does a notion like \locality. As has been stated a few times above, 
orthodox quantum mechanics respects \locality, even when it is understood as a realistic theory, 
an understanding held by most physicists who don't think long and hard about foundations (and by some who do). 
That is, even the process  
whereby, when Alice and Bob share a singlet state,  a measurement by Alice in a certain basis 
causes the quantum state of Bob's system to collapse instantaneously into one of the basis states, does not violate \locality. Yet the very wording of the preceding sentence implies that the described process {\em does} 
violate \relativity\footnote{Notably, the faster-than-light effect of Alice's choice on Bob's conditioned state has now 
been verified experimentally with no detector efficiency loophole~\cite{Wittman12}, unlike {\sc Bell-nonlocality}~\cite{Loopholes2014}.}. 
By contrast, operational quantum mechanics does not violate \relativity, because 
it does not entail any causal narrative involving quantum states, but simply uses them as computational 
tools. A more precise formulation of this idea will be given elsewhere~\cite{WisCav15}.

A similar example of the advantage of talking about causes is the ability to formulate \free\ in an obvious way: 
that a freely chosen action has no relevant {\sc causes}\footnote{\blk This may sound like a strong statement, and the reader may feel tempted to follow neither the operationalist nor the realist camp, but rather to reject Postulate~\ref{po:fc} from the list of assumptions in theorem~\ref{thm:rec}. This temptation should vanish if the reader thinks through what it would actually mean to explain away Bell-correlations through the 
real (not just in-principle) failure of \free. There is no general theory that does this. 
If such a theory did exist, it would require a grand conspiracy of causal relationships 
leading to results in precise agreement with quantum mechanics, even though the theory itself would 
bear no resemblance to quantum mechanics. Moreover, it is hard to imagine why it should only be 
in Bell experiments that free choices would be significantly influenced by causes relevant also to 
the observed outcomes; rather, every conclusion based upon observed correlations, scientific or casual, 
would be meaningless because the observers's method would always be suspect. It seems to us that 
any such theory would be about as plausible, and appealing, as, belief in ubiquitous alien mind-control.\label{foot:alien}}. 
% \egc{That is not at all obvious. Rather it's most likely false! :P}
This can be opposed to Colbeck and Renner's assumption {\em FR}, which also supposedly 
corresponds to ``the assumption that measurements can be chosen freely''~\cite{ColRen11}, 
but which is actually the assumption of \la\ as defined in Sec.~\ref{sec:repr}. 
As we have seen (Lemma~\ref{lem:la}), this can be derived from \free\ only by also using the Postulates of \relativity\ 
and \common\ (or something similar). The assumption {\em FR} clearly embodies far more
than merely freedom of choice, and interpreting it as if it does not (\ie\ taking their 
 language at face value) leads to bizarre conclusions. For example, Colbeck and Renner 
would be forced to claim that a typical physicist (see preceding paragraph), adhering to the realistic interpretation of orthodox quantum mechanics, {\em does not} believe in freedom of choice, 
{\em even if} said physicist were to believe that humans are not wholly governed by physical laws 
and have free will  in the strongest possible philosophical sense. 

Formalizing notions relating to Bell's theorem as causal principles also makes it apparent 
that some notions are more natural than others. In particular, the notions of \locality\ and 
\predet, which Bell introduced in his first paper, are not very natural. The first is weaker 
than the notion of \la\ that can be derived from the more fundamental postulates of \ac\ and \relativity, 
while the second is even more contrived. The latter criticism we would also level, even more strongly, at another 
notion which has been suggested as a replacement for \predet, namely `completeness'~\cite{Jar84}
(or `Jarrett-completeness' as one of us has called it~\cite{Wis14b}), also known as `outcome-independence'~\cite{Shi84}. 
In this context, more natural concepts for formulating a theorem in the style of 
Bell's 1964 theorem will be considered elsewhere~\cite{WisCav15}. \egc{Should we leave the title out, in case we change it later?}\hmw{ok, done.}

Finally, approaching Bell's theorem using ideas of causation can lead in new directions. 
For instance, {\em if} one assumes the predictions of relativistic quantum mechanics to be correct, 
%and that there are no Newcomb-predictors in our world~\cite{Eric-suggestion?}, \hmw{Eric, suggestion?} 
then it seems that 
one can, in Theorem~\ref{thm:rec}, replace the two Postulates 
of \free\ and \relativity\ by the single Postulate of {\sc no fine-tuning}~\cite{WooSpe15}. This 
is a postulate that the conditional independence relations between 
observable events are a consequence only of the causal structure 
(\ie~which events are {\sc causes} for which other events), and are not to be explained 
by fine-tuning of the probabilities of events. Whether this formulation can really be regarded as 
having fewer postulates than the above, whether it is truly possible to dispense with \free, 
and how the two camps could be expected to react to it,  are interesting questions for exploration.

%\begin{principle}[\JC]  
%For any observable event $A$, and any SLH, there exist %causes for an observable event [I SEE NO USE FOR THIS]
%events prior to that SLH, which, together with free choices 
%subsequent to that SLH, condition $A$ to be 
%independent of any observable events that are not its effects.% space-like-separated from $A$.
%\end{principle} 

\section*{Acknowledgements}
This research was supported by the ARC Discovery Project DP140100648 and ARC DECRA project DE120100559. 
We thank Eleanor Rieffel for helpful comments.

\bibliography{../QMCrefsPLUS}

\end{document}